# COVID-19 Remote Patient Monitoring: Social Impact of AI

Ashlesha Nesarikar[1]  Waqas Haque,  Suchith Vuppala,  Abhijit Nesarikar
*University of Texas at Dallas*  *UT Southwestern*  *UT Southwestern*  *Plano Intelligence*
*Plano Intelligence*

May 2020

## Abstract

A primary indicator of success in the fight against COVID-19 is avoiding stress on critical care infrastructure and services (CCIS). However, CCIS will likely remain stressed until sustained herd immunity is built, either in due course or by vaccination on mass scale. There are also secondary considerations for success: mitigating economic damage; curbing the spread of misinformation, improving morale, and preserving a sense of control; building global trust for diplomacy, trade and travel; and restoring reliability and normalcy to day-to-day life, among others. We envision technology plays a pivotal role. Here, we focus on the effective use of readily available technology to improve the primary and secondary success criteria for the fight against SARS-CoV-2. In a multifaceted technology approach, in this Part I, we start with effective technology use for remote patient monitoring (RPM) of COVID-19 with the following objectives:

1. Deploying readily available technology for continuous real-time remote monitoring of patient vitals with the help of biosensors on a large scale.
2. Effective and safe remote large-scale *communitywide care* (mass health monitoring with a cognitive AI platform) of low-severity cases as a buffer against surges in COVID-19 hospitalizations to reduce strain on critical care services and emergency hospitals.
3. Improving the patient, their family, and their community's sense of control and morale.
4. Proposing a clear technology and medical definition of remote patient monitoring for COVID-19 to address an urgent technology need; address obfuscated, narrow, and erroneous information and provide examples; and urge publishers to be clear and complete in their disclosures.
5. Leveraging the cloud-based distributed cognitive RPM platform for community leaders and decision makers to enable planning and resource management, pandemic research, damage prevention and containment, and receiving feedback on their strategies and executions. More details to follow on this in Part II.

---

[1] Corresponding author: ashlesha.nesarikar@utdallas.edu
URL: https://www.planointelligence.com

# I. Introduction

Social distancing and coordinated community action appear to slow the exponential increase in new COVID-19 cases[1, 2] for the first wave of the pandemic. However, COVID-19 poses pressing challenges not seen with recent pandemics[3, 4]. The mortality rate and contagiousness of COVID-19 is likely to be higher than that seen in the H1N1 pandemic in 2009[1, 2]. The current pandemic is also compared to the 1918 Pandemic Influenza, which occurred in three waves across the spring, fall, and winter of 1918[3, 5]. Moreover, the 1918 Pandemic Influenza spread over a year, while COVID-19 reached a global scope within weeks. As of May 15, 2020, earlier strain on critical care infrastructure and services (CCIS) appears reduced to some degree; on the other hand, the inherent weakness of an in-person-care (also referred to as brick-and-mortar) model is apparent. The current standard of care, financial incentive structure[6], and infrastructure relies heavily on brick-and-mortar based health services. In waiting rooms, patients and non-patients are vulnerable to potential exposure to COVID-19. Patients with other chronic illnesses and conditions susceptible to COVID-19 complications are especially at risk. COVID-19 has upended earlier best practices. The critical care system is especially strained, leaving many traditionally high-risk patients without care[6]. Moreover, the pandemic has pushed health services at all levels to strained conditions not seen before. In extreme cases, some communities have seen their health and social care system brought to their breaking point as localities run out of personal protective equipment (PPE), ICU beds, and ventilators, among others. For some communities, rationing health services may be unavoidable[7].

Even in the most optimistic scenario—if pandemic history is any guide—the worst may be yet to come; the crisis may continue for up to the next 18 months[1, 3, 4, 8]. It is time to utilize the success of first wave containment—the accumulated time, experience, and community spirit—to prepare for a likely second wave[9]. Technology has the capability and readiness to make a social impact not only within local communities, but globally. Raztan et al. report communication and data voids due to uncertainty around the trajectory of COVID-19; this uncertainly is one of the causes of fear and misinformation[9].

COVID-19 is evolving, and presents new, perplexing symptoms that are the subject of ongoing investigations[10, 11, 12]. Tang et al. found that blood coagulation abnormalities (increased D-dimer and fibrin degradation product) were markers for COVID-19 mortality[11]. Thachil et al. in their guidance for COVID-19 coagulopathy, report a 3-fold increase in D-dimer values (among other coagulopathy markers) indicates a significant need for hospital care even if there are no other concerns[12]. They recognize that their guidance will change with evolving knowledge. Such a rapidly changing disease, healthcare guidelines, and community response requires a modular and scalable system for remote patient monitoring (RPM). Our cognitive RPM system is up to the task as it is component- and cloud-based, and may be readily deployed in the cloud on demand.

We propose remote patient monitoring (RPM) on a mass-scale (*communitywide care*) as one way to prepare for possible subsequent waves of rapid increase in SARS-CoV-2 infections; we refer to these rapid increases in COVID-19 cases as *surges*. The primary objective of RPM is to counter the uncertainties associated with COVID-19 and its surges. We propose gathering community wide continuous real-time COVID-19 patient data for all patients. This data may help with understanding the *progression* of COVID-19 in individuals, the *spread* of it community wide, *containment* effectiveness for the community, and the community's *pandemic role* [9]. We recognize that progression, spread, containment, and pandemic role are topics of investigation in their own right. Our approach is to use a cloud-based AI platform (cognitive RPM system) to generate data and intelligence that may further the respective investigations. Besides the data and intelligence derived from it, we offer a unified and common source of data that may be cross referenced across different research disciplines related to progression, spread, containment, and pandemic role. The referential integrity and traceability of the data across disciplines, we believe, is one effective tool in combating surges and uncertainty surrounding COVID-19. In Part I, we primarily focus on progression and spread. We hope to learn from the community and peer feedback on Part I, which may help us cover containment and pandemic role in Part II. Our systems approach—in this Part I and the proposed Part II—to counter COVID-19 is an approach called for by Ratzan et al. [9] in being prepared and proactive in preventing, managing, and mitigating COVID-19 surges. The systems approach may not only subdue adversities against health, health workers, and health infrastructure, but also challenges (covered in Part II) faced by businesses, civic planning, economic forecasting, and broader societal responses to the pandemic.

## II. COVID-19 Symptoms and Severity

Remote patient monitoring (RPM) is proposed to care for patients with various diseases[13, 14, 15, 16, 17]. In this paper, we focus on COVID-19. We divide patients into two categories based on their symptoms[10]:

- Low-severity: Symptoms include fever, fatigue, myalgia, cough, sore throat, runny nose, sneezing, nausea, vomiting, abdominal pain, and diarrhea (these are associated with acute respiratory tract infection and digestive distress) [10]. We also include high-risk (due to key mortality markers including high blood pressure, diabetes, blood coagulation abnormalities, COPD, etc.) patients suspected to have or tested positive for COVID-19, but who show no symptoms.
- High-severity: Symptoms include pneumonia with hypoxemia, $SpO_2 < 92\%$. Acute respiratory distress syndrome (ARDS), shock, encephalopathy, myocardial injury, heart failure, coagulation dysfunction, and kidney injury[10]. We also include a three or more fold increase in D-dimer values based on current reports and guidance[11,12].

A low-severity COVID-19 patient may be referred for home based real-time monitoring with the cognitive RPM system. The AI-based cognitive RPM system monitors the patient for their propensity to a potential turn to high-severity. If high-severity COVID-19 is expected or detected, patients, doctors, paramedics, hospitals, resource managers, and other city officials may be notified. A patient may be transferred directly to a hospital bed.

## III. Real-time vs On Demand Monitoring

Real-time monitoring offers several benefits over on-demand monitoring. Real-time is used to mean many things in literature depending on context. In an engineering sense, a real-time observation of an event indicates an inconsequentially small time lag (e.g., to maximize resolution) between the event and its observation, while maintaining continuity of observation over the duration of the event. Traditionally, engineers have accepted a tradeoff between resolution and sampling continuity. On demand monitoring sacrifices both sampling continuity and resolution to allow human consumption of the raw data. If we take 12 days[18] as the progression duration of COVID-19, a typical real-time observation of the patient symptoms (e.g., with biosensors) may involve taking observations every second for the 12-day duration. A ten-minute reading of sensors on each of the ten days whether at home, online, or in a clinical setting, is not a real-time observation. Sporadic observation of symptoms over the 12-day period are considered on-demand observations.

An on-demand observation of an event that is valid over a given set of time intervals (representing fractions of the event's duration) is time dependent. On the other hand, a continuous real-time observation over the event's entire duration is time independent. Real-time and on-demand are context-sensitive concepts that depend on data variability of the process as a function of time. For a chronic disease like Type II diabetes, 12 evenly spaced blood sugar measurements in a day over a period of a month (e.g., the time between doctor's office visits) may be regarded as real-time, taking into account a patient's daily activities and metabolic/activity cycles in aggregate. On the other hand, COVID-19 is a fast progressing disease that is fraught with uncertainties; ideally, we may want to emulate data collection frequencies used in an ER environment and collect readings continuously for almost every second. The need for such continuous real-time readings is validated by reports of rapid onset of life-threatening conditions in previously symptom-free patients[10, 19]. Therefore, real-time data collection for COVID-19 may involve recording each relevant vital on the order of a second for about two weeks or until the infection is cleared. We name a set of relevant vitals along with its concurrently collected date-and-time stamp a *data-point*. A data-point is collected about every second, and it represents a snapshot of the patient's condition that may allow a doctor to arrive at a *diagnostic* with reasonable confidence. We use these definitions of real-time and data-point in the COVID-19 monitoring and diagnostic context.

Typically, physicians are accustomed to on-demand measurements; the average doctor may find it adequate to collect one sample of $SpO_2$ every hour to monitor a COVID-19 patient. Real-time data in its raw form may be overwhelming and counterproductive for the doctor. Therefore, systems that collect real-time data need to include modules for analysis, proactive intelligence generation, and visualization so that the doctor's productivity and effectiveness are improved. Real-time measurement and the large amount of data generated in the process is a prerequisite for machine learning and AI approaches to COVID-19 monitoring. Unlike a doctor, the AI of the RPM system proposed here can monitor COVID-19 patients continuously without fatigue, in large volumes, with short notice, and without consuming critical ER (CCIS) resources. Such a system can be part of the preparation for anticipated COVID-19 surges. In practical scenarios, (see Appendix B: COVID-19 Case Studies and Use Cases), real-time data gathering is subject to disruptions. A patient monitored by a real-time device may—knowingly or unknowingly—disconnect the device; the device may experience outages or misuse; or connectivity from the patient's home to the RPM server may suffer. These contingencies may be addressed technologically; moreover, similar contingencies may affect patients currently in ERs and hospitals. New challenges for such a real-time RPM system exist, and we address and overcome them with the cognitive AI platform.

1. Privacy: Patients and doctors alike should raise privacy concerns in any large scale monitoring system. As part of our cognitive AI system, we continue to comply with existing privacy laws and IRB approval (an independent expert review to ensure ethical methods).
2. Data deluge: Data collection for direct human consumption increases productivity only up to a certain level; after that, human productivity plummets due to excess data. We solve this challenge by generating cognitive intelligence from the data and provisioning the intelligence just in time, to the right individual (or collaborating system), in the right format, and to the extent desired by the individual. With cognitive RPM, a doctor responsible for oversight of a large number of COVID-19 patients would be notified only after certain high-severity alarm conditions are met. Then the doctor may ask for justification related to the alarm condition. Subsequently, the doctor may choose to remotely contact the patient in an on-demand session, or the doctor may direct paramedics to transport the patient to an ER.
3. False positives or other inundation of information: Not all AI systems are the same. Tactically collected sensor data is not enough; doctors want generated tactical AI consolidated with other known facts to deliver actionable intelligence—also referred to as cognitive intelligence. Performance considerations for the cognitive RPM system's alarm condition (transition from low- to high-severity symptoms) are provided in the Communitywide Care Performance Measurements section. The system may be configured to collect feedback from doctors and generate an evaluation of its own intelligence generation.
4. Security: The cognitive AI platform employs contemporary security (e.g., securing of data per HIPAA, PCI, or custom client needs). Security along with privacy and performance are a core part of the platform's design; they may be verified and improved during major release cycles.

## IV. RPM for COVID-19

We propose a generic use case for self-administered RPM using biosensors (see Biosensors for COVID-19 Remote Monitoring) based on published criteria for determining COVID-19 progression (see Diagnostics, Disease Progression, and Health Predictors) in real-time on the cognitive AI platform (see Appendix A: Real-Time Monitoring, Intelligence Generation, and Notification Distribution). We insist on real-time monitoring over on-demand monitoring due to uncertainties regarding COVID-19 progression, unexpected turn of symptoms to high-severity (e.g., due to hypoxia), and the potential inability of patients to notice a sudden turn for the worse; for example, silent hypoxia as reported by Levitan[19]. While high-severity patients need hospitalization and potentially ER monitoring, low-severity patients' use of hospital resources may exert undue strain on CCIS[3]. Real-time differentiation and identification of high-severity patients from a pool of COVID-19 cases in a community under surge is the differentiating benefit of our proposal and use of the cognitive AI platform. In our proposal, we:

1. Monitor *all* positively tested COVID-19 cases with low-severity indicators in a community.
2. Predict and identify—in real-time—inflection points from low- to high-severity; generate notifications of the inflection points as alarm conditions; deliver the notifications and the relevant supporting evidence to health-practitioners, paramedics, and authorized individuals; and make the relevant data and its analytics available on demand.
3. Support critical care infrastructure and services (CCIS) in its existing format (there may not be enough time to modify, test, and validate the already efficient functions of CCIS; e.g. ER procedures may already be optimal). For example, feedback from hospitals, ER professionals, and paramedics may be utilized to derive alarm conditions; deliver them as real-time notifications; and provision supporting data to justify the alarm conditions to the notification recipients.
4. Collect, analyze, and distribute the data—within regulatory compliance guidelines (e.g., HIPAA, security, role-based access, IRB review and its guidelines, FDA and CDC recommendations and guidelines, etc.)—to pandemic research; resource management; planning, prevention, and containment efforts; and community awareness and education efforts; among others. More details on this will be covered in Part II.

A. COVID-19 remote-healthcare: On-Demand vs. Real-Time

Both on-demand and real-time tools may be used for remote-health provisioning. The scope of this paper is limited to remote monitoring of patients experiencing symptoms of low-severity COVID-19. A patient diagnosed with or suspected to have worsening hypoxia or other medical complications needs care in a clinical or hospital setting. For remote monitoring of low-severity COVID-19 patients, we describe two approaches to remote-healthcare and their cost benefits: Approach I is on-demand remote monitoring of patients by their doctor, and Approach II is real-time monitoring of patients by an AI system.

B. Approach I: On-demand monitoring

On-demand monitoring is widely covered in the literature and this is the only option currently available for remote monitoring. Real-time monitoring is not addressed in otherwise comprehensive guidelines of on-demand telehealth provided by HHS[20]. Some on-demand monitoring solutions are referred or implied in their disclosures to be real-time, though in reality, they are at best episodic-care approaches. At a scheduled time—at the request of the patient or the doctor—a doctor on a video or audio conference with a patient may gather symptoms and results of patient self-administered tests. The patient may use a measurement device at the discretion of the doctor to gather health data. Successful implementation, cost-benefit, and challenges of on-demand monitoring are subject to:

a) The patient's ability to follow the doctor's directions in gathering diagnostic data. Typically, a doctor's ability may be limited in guiding a patient to collect, note, and convey a meaningful data sample during the on-demand session. Several studies have reported varying degrees of success rates[13, 14, 15, 16, 17]. Patient training, ease of use of devices, intuitiveness of interfaces, and in-home support for the patients may be influencing factors for success of on-demand monitoring.
b) Rapid symptom progression. For example, in the case of COVID-19, on-demand monitoring may miss rapid progression to high-severity symptoms. The doctor may primarily rely on and be limited by the episodic data collection strategy of the on-demand session for diagnosis; potential red flags may be delayed or even missed. The success of this approach is primarily dependent on cooperation, communication, and joint understanding between the patient and doctor.
c) User familiarity with the system. On-demand monitoring is a natural extension of medicine as it is practiced today. A phone call or a video call that the patients and the doctors are already familiar with may not differ significantly from on-demand monitoring (for example, a doctor asking for a patient to take his temperature during a video call). In real-time AI monitoring, the possible involvement of a potentially opaque system—for real-time data collection, data analysis, and decision making—may be perceived by both the patients and the doctors as novel and may need some getting used to.
d) Availability and scalability. On-demand solutions and their technology variations are readily available from several vendors. Though on-demand monitoring is promoted as a solution to remote monitoring of COVID-19, it is neither real-time continuous monitoring nor scalable. Patients may not be monitored continuously, and at any one time, only one patient can be monitored by one health care professional.

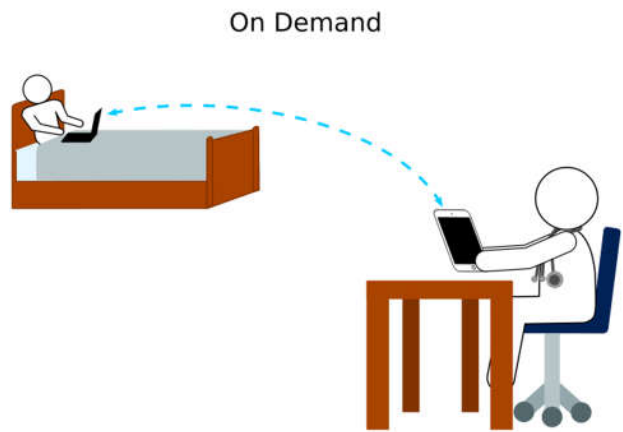

*On-demand—a dedicated limited-time communication link. An example of episodic monitoring as opposed to real-time monitoring.*

### C. Approach II: Real-time monitoring

Continuous real-time patient monitoring by scalable, distributed, artificial intelligence based systems are, to the best of our knowledge, not in use or otherwise proposed in the literature. Based on reported COVID-19 spread[2, 5, 8, 10] characteristics, we envision a need for mass-monitoring of vitals that can be scaled up on short notice. Most positive cases are expected to recover without leading to high-severity or life-threatening symptoms[2, 5, 10]; however, almost all low-severity cases would need continuous monitoring on a large scale (also referred to as communitywide care). Though aggravation of symptoms may be sudden and unexpected, only a small fraction of COVID-19 cases is expected to exhibit life threatening conditions and need for ER care. As an illustration, we present typical communitywide care scenarios (see Appendix B: COVID-19 Case Studies and Use Cases) with the platform. We draw on our previous experience (see Appendix A: Real-Time Monitoring, Intelligence Generation, and Notification Distribution) in our proposal to design, implement, and test real-time data collection, intelligent screening, and notification in combating COVID-19.

For the purposes of this paper, we consider home-based vitals monitoring scenarios (see Appendix B: COVID-19 Case Studies and Use Cases). This RPM scenario may be extended to emergency non-hospital quarantine locations in the event of a rapid surge in positive COVID-19 cases. We identify typical vitals, their thresholds, and alarm conditions that need monitoring, emphasizing rapid identification of patients with deteriorating conditions. Our objective is to identify patients that may need hospitalization and potentially ER care before their condition deteriorates to high severity. The success of communitywide care may be measured in terms of false positives and false negatives (see Communitywide Care Performance Measurements).

Successful implementation of real-time monitoring, especially as compared to on-demand monitoring, depends on:

a) Ability to emulate ER-like (continuous real-time) monitoring in non-hospital settings. For our home-based monitoring scenario, we need to collect data frequently and continuously. In this context, we differentiate the cognitive RPM system from a generic continuous data collection system with smart phone app, graphs, and numerical displays. Our system can generate cognitive intelligence and deliver it just in time, to the right end users, in a format suited to the end user, with the extent of detail (e.g., metadata) specified by the end user. The deluge of information generated by the generic app is often detrimental to the cognitive productivity of the end user (e.g., doctor responsible for overseeing RPM for several patients). As an example, for a multi-tasking doctor, the cognitive RPM system shields the doctor from the data deluge, generates actionable intelligence, and delivers it to enhance the doctor's cognitive productivity, induce intelligent insights, and reduce multi-tasking fatigue.

b) Real-time analysis of and intelligence generation from the collected data to project the need for patient hospitalization to prevent life threatening conditions. Our initial approach is for the cognitive RPM system to identify alarm conditions and generate and deliver real-time notification of the alarm conditions (e.g., to PCP, ER physicians, or paramedics) and, if needed, patient vitals history and PCP (primary care physician) notes.

We expect the availability of a patient's vitals history and PCP notes at the time of ER admittance will reduce wait, processing and diagnosis time.
c) Shifting COVID-19 patient screening burden from ERs and hospital waiting rooms to patient homes (or other designated quarantine locations). The patient's PCP may monitor non-emergency progression of the disease or evaluate the need for hospitalization. This may also allow PCPs to see their non-COVID-19 patients in their regular practice.
d) Bringing greater predictability to the progression of COVID-19 cases in a community by allowing the community and its administrators to monitor the progression of COVID-19 in real-time, generate high-confidence projections, manage resources effectively, and take proactive steps to prevent expected future adversities. Details of this functionality will be provided in Part II.

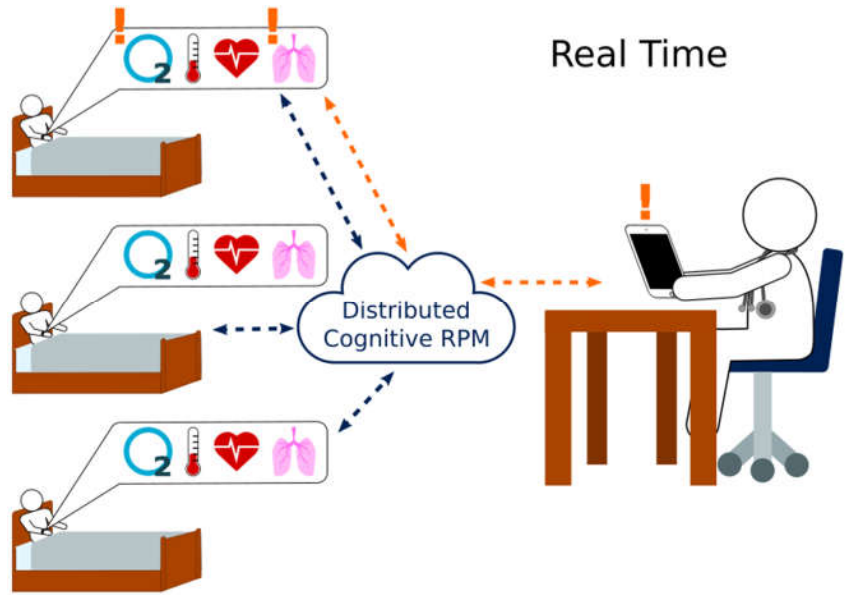

*Continuous real-time patient monitoring by a scalable, distributed, AI-based system. Multiple patients can be accommodated simultaneously. The doctor is proactively alerted to patients who may require attention.*

## V. Diagnostics, Disease Progression, and Health Predictors

Rapid progression of COVID-19 and the resulting unprecedented pandemic motivated us to compile symptoms, severity and morbidity indicators, and related health features of the disease. The information in this paper is not health advice, a diagnostic indicator, or patient guidance.

A patient's COVID-19 health predictors may be divided into two categories: the patient's vitals and the patient's medical history. Vitals are measured in real-time to track progression and pinpoint potential deterioration to high-severity symptoms. Real-time measurements of vitals is at the core of the cognitive RPM system. We propose a list of vitals that may need monitoring for effective real-time RPM.

A. Patient Vitals
1. Respiratory Rate: This is the number of breaths per minute. A normal range for adults is 12 to 20 breaths per minute, with a lower number for individuals with better cardiovascular fitness. During COVID-19 progression, an increased respiratory rate over a short period of time may indicate a transition from low-severity to high-severity; this may be accompanied by shortness of breath. Breathlessness typically develops between day 5[21] and 8[22] of symptoms. A review of risk factors for progression of COVID-19 found that shortness of breath (dyspnea) is more significantly associated with mortality than presence of a fever[23]; dyspnea was reported in between 40 and 87% of patients[23]. Note that our objective is to generate an alarm condition in the cognitive RPM system well before the patient develops breathlessness or silent respiratory distress (such as the occurrence of silent hypoxia[19])
2. Blood Oxygen Saturation ($SpO_2$): This measures the percentage of oxygen saturation in the blood. Normal range is 95 to 100, and an alarm condition may be met for levels persistently below 92. This is likely the most important vital for COVID-19 patients.
3. Heart Rate: Elevated heart rate (above 100 beats per minute) is believed to be one of the signs of worsening condition. However, no published research linking elevated heart rate to COVID-19 was found.
4. Blood Pressure: Numerous studies have found that patients diagnosed with hypertension have an increased risk of severe COVID-19 presentation and mortality[24]
5. Temperature: Shi et al. report lack of positive correlation between fever and severity of COVID-19[23].

A patient's continuous self-assessment of clinical symptoms is an important tool for determining alarm condition. This is a subjective measure of symptom severity that depends on patient characteristics, behavior, and daily routine, along with their evaluation by the patient's physician.

B. Patient History

Patient history guides predictors of mortality, propensity for high-severity symptoms, and expectations of outcomes in general. The current understanding of severity and mortality predictors may change due to uncertainty and lack of evidence-based guidelines from peer reviewed data on COVID-19. Currently, long-term effects of severe COVID-19 presentation or resulting damage to lungs or other organs are not known. We propose to account for comorbidities[24] (e.g., diabetes, hypertension, COPD or other respiratory diseases, underlying vascular disease, etc.) and other adverse factors (e.g., smoking, tobacco use, age over 65, history of lung or other cancer, etc.) from patient history in formulating an RPM alarm condition in Part I. We plan to cover community impact, use of data (e.g., for research, planning, communication, education, etc.), and potential use of the cognitive RPM system for studying the COVID-19 pandemic in Part II.

## VI. Biosensors for COVID-19 Remote Monitoring

Various telehealth and on-demand remote monitoring proposals have been reported—for example, RPM for management of chronic diseases[14] including COPD[15, 16] and diabetes[6], and RPM as a part of mobile healthcare in rural and remote regions[13]. Though a wide variety of devices are reported, not all devices are suitable for real-time monitoring; reporting on usability, efficacy, and reliability is in its early stages[13]. On the other hand, consumer wearables to self-monitor and track personal fitness and health habits have been successful over the last decade. In the US, one in six fitness and health enthusiasts have used at least one device to self-monitor with a smart phone or smart watch[26]. Unlike clinical devices, consumer devices may lack key requirements for real-time RPM, including complete disclosure of technical specifications and supporting medical and efficacy data.

Unlike for consumer devices, the rapid response needed for COVID-19 remote monitoring needs to account for accuracy and reliability of the self-administered devices. Consider, for example, oximeters as essential devices for monitoring COVID-19. We find that the marketplace lacks a clear demarcation between clinical and consumer grade oximeters. PCP and ER doctors need to rely on the real-time and historic data to make potentially critical patient health decisions. A myriad of cheap devices is available from global suppliers that are not subject to US regulatory supervision or approval. In 2016, Lipnick et al. [25] reported four of six considered oximeters have subpar accuracy and reliability. For hypoxia scenarios, four of the six reported errors for blood oxygen saturation ($SpO_2$) levels $< 80\%$ and three of the six had errors for $SpO_2$ levels between $80 - 90\%$. Even FDA approved devices for use in clinical settings may not meet the World Federation of Societies of Anaesthesiologists standards[25].

Self-administration of remote devices may pose further challenges and is subject to erroneous readings due to an oximeter on a patient's finger being susceptible to: hypoperfusion of the finger and its hand that is unrelated to patient health (e.g., reduction in natural blood flow due to the hand being relatively cold or due to vasoconstriction); the patient's movements; and calloused skin on the finger. Erroneous readings might also result from inability to follow instructions (e.g., in application, reading, and maintenance of the oximeter); operation of the oximeter outside its design specification (e.g., subjecting the oximeter to excess temperature or excess pressure—for example, by sleeping on it); and unexpected patient activity (e.g., physical exertion of a patient who is prescribed bed rest). We include the errors and potential misadministration of devices as a key factor in measuring effectiveness and efficacy of the cognitive RPM system (see Communitywide Care Performance Measurements).

The cognitive AI platform is designed to generate intelligence from multiple IoT devices. It can consume data from scientific (e.g., clinical grade for COVID-19 monitoring), commercial wearable, and commodity AV (audio video) devices concurrently to derive actionable intelligence. Cognitive intelligence generated by the platform may compensate for heterogeneity of IoT devices, an erroneous device reading, and patient-induced anomalies in measurements.

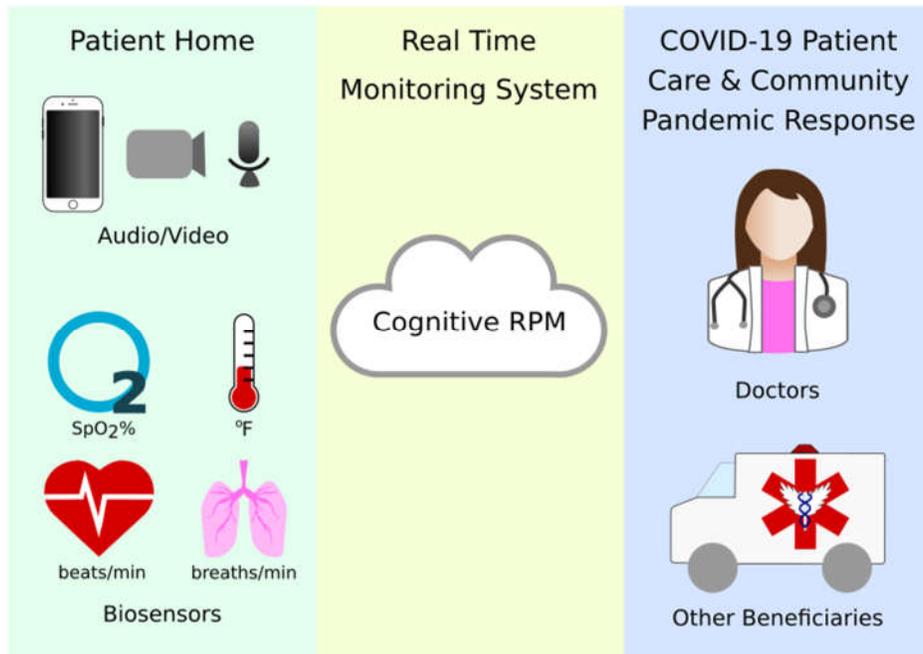

*Cognitive RPM architecture: the cloud-based cognitive AI system can process various data from multiple devices and derive intelligence for the use of the PCP and other COVID-19 pandemic response.*

## VII. Communitywide Care Performance Measurements

The cognitive RPM system and its servers may be evaluated with functional and non-functional performance measures. Functional performance measures achievement of application-specific functionality. Non-functional performance accounts for general efficiencies and compliance of the platform. Appendix A: Real-Time Monitoring, Intelligence Generation, and Notification Distribution describes the core cognitive AI platform and its key non-functional performance indicators including bandwidth, CPU vs GPU usage, and computability in general.

Here, we focus on functional performance measures for remote monitoring of COVID-19 patients in their homes. While cognitive RPM functionality may be greatly influenced by the community, its jurisdictions, and its doctors, we present functional performance measures for a generic use case (see Appendix B: COVID-19 Case Studies and Use Cases). The functionality represents intelligent abilities (a human equivalent) of the cognitive RPM system to meet target expectations. Three target tasks for the system are:

1. Detect patient non-compliance: This is regarded as a challenge to real-time monitoring that may be overcome by patient training and regular (remote) oversight by health professionals. Patient non-compliance includes:
    a. Intentional or unintentional removal of sensors. For example, the system needs to detect removal of the sensors by the patient for a shower, bathroom break, unintentional removal of sensors, etc.
    b. Performing activities (e.g., exercise) which may alter sensor data against the advice of the doctor.

2. Detect faults in monitoring devices: Monitoring devices may generate faulty data or stop entirely due to:
    a. Low battery or lack or power.
    b. Adverse environmental conditions, e.g., high temperature, humidity, water immersion, etc.
    c. Loss of optimal device calibration.
3. Diagnostics for COVID-19 symptoms: With patient compliance and fault-free devices, cognitive AI can process the generated data-points to deliver intelligent diagnostics and an alarm (if warranted based on alarm conditions). Accuracy and time-to-delivery are important performance measures.

A. Performance Criteria for Project Execution and Delivery

Surges (of COVID-19 cases) are expected to continue into the next year[1,3,4]. Rapid deployment of COVID-19 remote monitoring depends on several civic, healthcare, organizational, and technical criteria; some of the important ones are listed below:

1. <u>Multidisciplinary Execution Synergy</u>: Rapid deployment and efficient execution may depend heavily on team members and stakeholders from healthcare, first responders, public policy, jurisdictional governance, civic planning, and technology. Execution plans need to account for efficient communication, go-no-go and milestones, and resource allocation, among others. A fast and effective delivery hinges on proactive initiative and participation from all areas of the team.
2. <u>User Adoption</u>: User acceptance is key. Regardless of a jurisdictional mandate, a broader and informed community adoption of the technology solution hinges on training and familiarity. Frequent user feedback may dictate the efficacy of a remote monitoring solution. Findings, successes, and lessons learned need to be published in reproducible detail.
3. <u>Findings Transparency:</u> Findings and lessons learned need to be published with clear definitions of success criteria, the extent to which the success criteria were met, and lessons learned. Urgency for a COVID-19 solution based on a scientific approach should not compromise reporting quality and detail; a political declaration in a research publication[17] is counterproductive.
4. <u>Thoroughness of Disclosure:</u> Promoters of COVID-19 technologies need to be aware of the unusual and urgent nature of the need. Devices and monitoring technologies implied to be COVID-19 ready need to meet the minimum criteria for technology readiness; e.g., clinical accuracy of devices need to be clearly stated and readily available[25]. A device that does not reliably generate continuous data streams in real-time is not (and should not be promoted as) ready for real-time time COVID-19 monitoring.
5. <u>Execution Details</u>: Rapid planning, resource procurement, and execution are challenges for COVID-19 monitoring in their own right. Project management experts may have this capability; however, they need to go beyond high-level objectives and goals and work out the finer project details; they need to expect and effectively execute pivots where needed.
   6. <u>Funding Criteria:</u>
       a. Project Funding: Traditional project funding sources need to look beyond "safe bets" but well short of "moon shots". The execution plan and its capability are keys. Capable teams need to be proactively reached out to, and approval cycles need to be shortened.
       b. Health services payments: Healthcare systems are intertwined with a variety of payment models. Though a simple fee for service model is typically envisioned for remote monitoring services, challenges to other payments models remain[6, 27]. For digital services—and advanced remote monitoring—needs exist for reimbursement models, regulatory alignment, and feedback or evaluation criteria for efficacy and evaluation, among others.
   7. <u>Uniqueness of Community Need:</u> Frontline communities and their leaders are already on their toes. They need to participate early and guide the effort to maintain the rapid pace of delivery for their own communities. Every community is unique and needs unique adaptation of a common execution plan.

# VIII.     Conclusions

COVID-19, as a highly contagious disease with a prolonged incubation period and possibility of sudden turn to life threatening conditions, has challenged the traditional healthcare approach. Currently, patients self-assess symptom severity and go to a hospital or ER waiting room for help. Even with increased testing, patient self-monitoring, self-assessment, and initiative are determining factors for visiting a hospital or ER waiting room. The cognitive RPM system addresses some of the key shortfalls of the traditional approach, which include:

   1. In brick-and-mortar (traditional) COVID-19 care, the time between the onset of high-severity symptoms and a patient's reporting to an ER, waiting for service, diagnosis, and eventual admission to the ER may span hours or longer; this time may be critical to COVID-19 care and recovery. Cognitive RPM is designed to identify the first signs of high-severity symptoms and generate supporting diagnostic data for hospitals or ERs so that patients may avoid the waiting room.
   2. Waiting rooms are overwhelmed with patients with COVID-19 symptoms, causing stress on the health system, spread of COVID-19, and delayed care for patients in critical condition, among others. Cognitive RPM may significantly reduce head count and time-to-care in waiting rooms; our goal is to eliminate COVID-19 cases from waiting rooms to avoid waiting room related pandemic spread.
   3. COVID-19 patient anxiety is centered on inability to get PCP care, self-monitor the disease, and stay informed of the progression. Cognitive RPM allows PCP to get involved in the early or low-severity symptoms stage of COVID-19 without disrupting care to non-COVID-19 patients. The patients informed by cognitive AI notifications and their own PCP may be more likely to follow social distancing and other community guidelines to counter the pandemic. We plan to cover the wide effects of cognitive RPM on community pandemic response, healthcare planning, and ER resource management in Part II.
4. Rural and sparsely populated communities where hospitals may not be nearby to handle COVID-19 surges face particular hardship in traditional healthcare settings. Cognitive RPM may provide not only in-home monitoring, but also PCP care from doctors in urban areas or available doctors in geographically distant communities. The ability to anticipate and plan for COVID-19 ER cases may reduce healthcare, economic, and resource strain on rural communities.

Challenges to rapid COVID-19 pandemic community response include difficulty in rapidly forming multidisciplinary teams of doctors, community leaders, and locally available technologists; lack of readily available funding; transparency and completeness of disclosures on devices, projects, and community initiatives; inability to address unique needs of individual communities; lack of execution guidelines and historic precedent; and slow pace of user adoption for pandemic related policies, technology solutions, and altered way of life due to the pandemic. We tackle some of these challenges in designing cognitive RPM, bringing together a multidisciplinary team, community involvement, and disclosure of information. Some others—deploying and testing to create execution guidelines and assessing user adoption—remain due to their dependence on funding availability.

## IX. Comments and Disclosures

The contributors and authors of this publication contributed their time and resources to this publication without any financial compensation. The cognitive RPM system proposed here and other systems built using information in this paper are subject to jurisdictional approvals (e.g., regarding security, privacy, HIPAA, IRB review, etc.), and consent from patients and their doctors. Our hope is to quickly bring community leaders, technologists, academics, and business leaders up to speed. For example, technologists, system builders, and developers may get a quick start in building prototypes, demonstrations, and POCs (proof of concept) from this publication.

## X. Acknowledgments

We would like to thank Ann-Cathrin Guertler for providing editing assistance and comments throughout the process, and Anika Nesarikar for initial concept discussions and subsequent conversion of the concepts to diagrams.

# Appendix A: Real-Time Monitoring, Intelligence Generation, and Notification Distribution

Cognitive RPM is supported by a privacy and security centric cognitive AI platform. It derives intelligence from deluges of real-time data collected from many heterogeneous IoT devices. The intelligence is delivered in real-time to the intended recipients, in their suitable format, exactly when needed. The system is a reusable, plug & play, versatile foundation for existing and future AI systems (for example, see the public safety application below). We bring three different innovations together to construct the cognitive AI platform.

- It is a distributed, redundant, scalable, and fault tolerant pluggable framework of servers, networks, and bi-directional real-time asynchronous communication. Backward compatibility considerably simplifies interfacing with standards-based third-party systems regardless of their age. Our standards-based (e.g., XML and JSON) framework will allow us to grow without significant redesign, keeping up with future technology changes.
- A variety of AI (as well as filter, analytics, and probabilistic) components may be plugged into the framework to achieve a range of design objectives. We have had success with deep neural networks (DNNs) optimized for specific tasks.
- A suite of UI allows easy consumption of the generated intelligence and other messages. UI designs for all device form factors (smart phones, tablets, and desktops), major operating systems, and system types are ready. Further UI work is intended to bring more uniformity and intuitiveness to the system.

Users can interface with AI systems supported by the cognitive AI platform through its Command and Control Console (C&CC). The C&CC is a real time cloud-based interface for administrators or a dashboard for decision makers. It presents a unified view of relevant media and aggregated intelligence. Users can collaborate remotely through the console from anywhere in the world in real time (a low latency high-bandwidth underlying infrastructure is assumed). Collaborators may form groups and teams on the fly to consume, exchange, and deliver their expertise, which may be shared or added to the primary intelligence stream. It is secured by encryption, authentication, and authorization.

We created a public safety application as a pluggable component of the cognitive AI platform. Among its numerous features is the ability to detect and locate wielded firearms. It achieves this by real-time monitoring of multiple video feeds simultaneously. Upon detection of a wielded firearm, the system immediately sends authorized end users notifications or alarms containing information needed to respond to the event. Notable advantages over traditional video surveillance include 1) live video analysis (with critical event information automatically reaching recipients within seconds, if not less) over a reactive approach, and 2) scalable automated monitoring of multiple feeds simultaneously (whereas humans lose monitoring ability with each additional screen).

In a test of the prototype, we utilized a high-bandwidth (10 Gbps locally and 1 Gbps between datacenters) and low-latency (1 to 10 ms) geographically distributed network between Texas and Illinois.  Ping time between the server and the C&CC console was about 150 ms and reflected instantaneous network performance. In a public demonstration, we used two simultaneous live video streams, with a video processing backend to generate person, age, gender, and gun detection signals.  In the demonstration, we were restricted to the use of available CPU servers (32-CPU, 96 G Mem Ubuntu 16.04 LTS), which as expected, were the rate limiting steps. Use of CPUs (as opposed to GPUs), required reducing the video quality to 720p from 2K.  Adjustments to framerate, consistency of network status, and uniformity of server load proved to be critical factors to maintaining the steady state of the process. We maintained the steady state for about a month with average end-to-end system processing (SP) time between a wielded gun appearing in front of the camera to a notification of it appearing in the geographically separated admin console of about 400 ms. The gun notification was delivered to the end-user on their own handheld devices within about 450 ms.  In the public demonstration, the application exceeded the design goal, which was to send gun notifications within seconds.

We conducted tests using mid to low-end GPUs for AI processing, which reduced the SP time to under 30 ms; notifications were delivered in less than 100 ms.  We anticipate that use of high-end GPUs may eliminate AI processing as the rate-limiting step in its entirety.  We also successfully tested concurrent processing using multiple AI back ends.

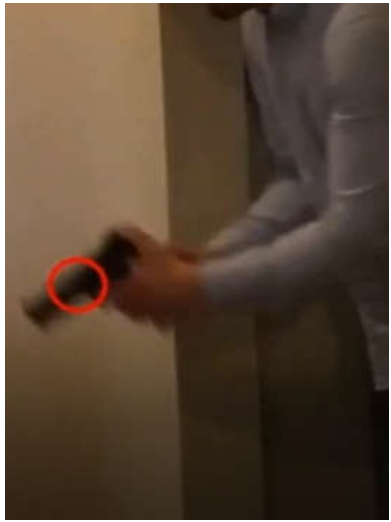

AI-based gun detection for a public safety application supported by our cognitive AI platform.

# Appendix B: COVID-19 RPM Case Studies and Use Cases

We use the following example case studies as basis for cognitive RPM use cases. The case studies are intended to provide sample scenarios of COVID-19 remote monitoring for the benefit of non-medical collaborators—for example, community leaders and technologists. The diversity of and details in the case studies indicate a need for cognitive AI over tactical AI.

Case Example 1:

A doctor sees a 60-year-old man with chronic obstructive pulmonary disease (COPD) with a long history of smoking, diabetes and hypertension. This individual has had COPD exacerbations requiring hospitalization in the past, and tested positive for COVID-19. After three days of in-hospital observation, the patient's clinical status was stable (respiratory rate of 12 with an oxygen saturation of 95 on room air). The patient wanted to go home on the condition he would continue to self-isolate. Hospital staff educated and sent the man home with cognitive RPM technology. Since the patient has multiple health conditions, the physician set a lower alarm threshold for the patient's oxygen status; the physician set the trigger at below 95% SpO2 instead of below 92% as for his younger patients without other significant health conditions.

Case Example 2:

A 30-year-old patient seeing a psychiatrist for a recent diagnosis of anxiety and history of panic attacks tests COVID-19 positive. The patient agrees to use cognitive RPM. On the third day of monitoring, the patient's vitals become unstable (heart rate rises to 130 beats per minute, blood pressure of 150/90, and respiratory rate of 35 breaths per minute) for 5 minutes before returning to normal. This triggers the alarm and the physician is notified. The physician reviews the transient episode (associated data-points, intelligence, and metadata), calls the patient, confirms the patient was having an anxiety attack, and reassures the patient. This helped the patient avoid an emergency department visit.

Case Example 3:

A 75-year-old woman with heart failure with increasing work of breathing tests COVID-19 positive and is monitored from home. The cognitive RPM technology alerted the doctor, who notified the ER and advised the patient to get right to the emergency department when her oxygen status began to gradually deteriorate. This prevented her having to wait in the hospital waiting room, and from potentially infecting individuals around her. She received medical care in the intensive care unit and was given a ventilator. An effective triage policy, social distancing in the community, and the RPM technology allowed the hospital to keep a steady supply of ventilators available.